\documentclass[aps,pre,reprint,superscriptaddress,10pt]{revtex4-1}
\usepackage{amsmath, amssymb, graphicx, color, braket,mathtools}
\usepackage{bbold}
\usepackage{verbatim}

\begin{document}

\title{Rigorous Wilsonian Renormalization Group for impurity models with a spectral gap}

\author{Peter Zalom}
\email{zalomp@fzu.cz}
\affiliation{Institute of Physics, Czech Academy of Sciences, Na Slovance 2, CZ-18200 Praha 8, Czech Republic}

\date{\today}

\begin{abstract}
	
The Anderson impurity model (AIM) has long served as a cornerstone in the study of correlated electron systems. While numerical renormalization group (RG) offers great flexibility for metallic reservoirs, it becomes impossible in an unbiased way when a spectral gap $\Delta$ opens up in the tunneling density of states. The only known exception is provided by the superconducting bath. In this paper, we lift these limitations by a novel numerical RG procedure that employs a discretization of the gapped tunneling densities of states into patches which accumulate at the gap edges. This reveals an unusual double scaling which is a shared behavior by the superconducting and the scalar gapped AIMs. Moreover, it requires a special iterative diagonalization procedure with an alternating scheme for discarding states only every second iteration. The discretization and the diagonalization scheme form together, what we refer to as, the log-gap numerical RG. It is successfully applied to the superconducting and to the scalar gapped AIM. Consequently, it reveals that both models belong to the same RG equivalence class which manifests physically in common singlet-doublet quantum phase transitions accompanied by in-gap bound states of given parities. While superconducting AIM is mainly used for benchmarking of the log-gap numerical RG, we also rigorously confirm the phenomenon of in-gap states escaping into the continuum, which was recently indirectly considered in Ref.~\cite{Pokorny-2023}. The gapped AIM is then tackled in a first ever exact numerical RG approach and confirms quantitatively assertions based on models with auxiliary metallic leads \cite{Diniz-2020,Zalom-2021,Zalom-2021r,Zalom-2022}. Moreover, it reveals that calculations performed in Refs.~\cite{Chen-1998, Moca-2010, Pinto-2015} are of strictly approximate nature.
\end{abstract}


\maketitle

\section{Introduction \label{sec_intro}} 

The Anderson impurity model (AIM), proposed by Philip W. Anderson in 1961 \cite{Anderson-1961}, has proven to be a fundamental framework for understanding a wide range of phenomena, including heavy fermions \cite{Lee-1986,Hewson-book-1993}, quantum dots (QDs) in Coulomb blockade \cite{Fulton-1987,Cronenwett-1998,Pustilnik-2004,Mitchell-2017} or add-atoms on normal surfaces \cite{Madhavan-1998,Li-1998}. It describes the behavior of localized electrons interacting with a continuum of itinerant states, encapsulating the complex interplay between localized and delocalized degrees of freedom which is beyond the reach of perturbation theory.

The theoretical efforts it stimulated, culminated by the computer aided solution via the Wilsonian Renormalization Group (RG) \cite{Wilson-1975, Krishna-1980a, Krishna-1980b}, which is nowadays commonly known as the Numerical RG (NRG). Quite recently, Wilsonian RG has been also applied to the AIM in the analytic framework of the Functional RG techniques \cite{Kopietz-2010,Kopietz-2013}. We also stress that several non-perturbative but approximate approaches are known to capture some of the aspects of the AIM \cite{Hewson-1993, Janis-2007, Janis-2019}, while at finite temperatures, the purely numeric approach via the Quantum Monte Carlo (QMC) method is routinely used \cite{Takegahara-1992, Gull-2011, Gubernatis-2016}. 

The main obstacle the RG targets specifically, lies in the formation of mutually interconnected scales in the impurity systems. For metallic itinerant states, as the cascade approaches the Fermi energy, it eventually gives rise to the Kondo effect. However, just by opening a spectral gap $\Delta$, as present in superconducting \cite{Hecht-2008,Heinrich-2018,Meden-2019,Zalom-2021r,Zalom-2022} or semiconducting materials \cite{Diniz-2020, Galpin-2008,Chen-1998, Moca-2010, Zalom-2022}, we deprive the system of delocalized degrees of freedom at the Fermi energy which severely impedes the energy cascade.

Intuitively, one expects an ordinary scaling of metallic AIM at temperatures much larger then $\Delta$ (all energy scales in this paper are measured in units of bandwidth $2B$) to be present at least approximately. Formation of local magnetic moments and even Kondo-like screening is then anticipated for temperatures remaining much larger than the Kondo temperature $T_K$ obtained for $\Delta=0$. Contrary, at temperatures much smaller than $\Delta$ no states are available to screen the impurity and dramatic changes are expected. However, so-far our understanding is complete only for the superconducting bath \cite{Moca-2021}. As we show here, this is also due to the fortunate nature of the one-lead problem, where the ordinary scaling of the AIM with constant metallic TDOS is preserved. 

For general gapped AIMs, no unbiased methods have so-far reached the region of $\Delta \gg T$. While for QMC the restriction is fundamentally due to the computational resources, in the case of NRG the limitations are of methodological character as problems arise already in the first step when Wannier-like states are constructed for the delocalized electrons. Also, as shown here, the low temperature scaling turns out to be unusual and requires crucial modifications to the diagonalization step of NRG.

Up to date, general gapped AIMs are solved only indirectly or approximately with either a small but non-zero TDOS induced into the gap region by adding a weakly coupled metallic lead to the problem \cite{Zitko-Mott-2013, Diniz-2020,Zalom-2021,Zalom-2021r,Zalom-2022} or Wilson chains were constructed artificially from the corresponding models at $\Delta=0$  \cite{Chen-1998,Moca-2010,Pinto-2015}. While the first approach utilizes rigorous Wilsonian RG on a modified system, in the second approach, truncated Wilsonian chains of the AIM with constant metallic tunneling density of states (TDOS) are postulated to represent the gapped problem up to the energy scale of $\Delta$. The two-scaled nature of the gapped problem is thus completely missed. Such results, as also clearly demonstrated in this paper, should therefore be understood as an approximate attempt. Nevertheless, qualitative conclusions from all currently available methods consensually confirm an unscreened impurity in the doublet ground state (GS) at $T=0$ for completely particle-hole symmetric scenario. Quantum phase transitions (QPTs) from a doublet to a singlet GS are observed upon further changing the orbital filling or particle-hole symmetry of the band \cite{Moca-2010,Zalom-2022}.

While similar QPTs constitute basic phenomena in hosts of superconducting nature, the theoretic connection to the isolator/semiconductor problem was recognized only recently in Ref.~\cite{Zalom-2021}, where the superconducting AIM (SC-AIM) was mapped onto a model with scalar gapped TDOS. Nevertheless, the two-scaled nature was not deciphered as the resulting gapped AIM with specific TDOS was beyond the available NRG techniques and the system was augmented with a weakly coupled metallic lead. Additional studies of broad classes of gapped TDOS functions in Ref.~\cite{Zalom-2022} only reaffirmed these findings on a more general footing, but the study also suffered from implementing a weakly coupled metallic lead. Consequently, the existence of singlet-doublet QPTs could be so-far asserted only indirectly with the $T=0$ behavior of truly gapped systems being only extrapolated.

In this paper, we therefore develop NRG technique for general gapped AIMs with the afore-discussed two-scaled nature. We choose the gapped AIM with constant TDOS (defined in Sec.~\ref{sec_gapped_aim}) and the SC-AIM (defined in Sec.~\ref{sec_scaim_theory}) for the demonstrations. General restrictions on band discretizations are first given in Sec.~\ref{sec_gapped_tdos} with a general-purpose discretization proposed in Sec.~\ref{sec_loggap_disc}. The resulting Wilson chains are then finally revealing a two-scaled behavior which is qualitatively the same for both models despite additional divergences appearing in SC-AIM. The kept or discarded scheme of the standard NRG is then modified accordingly in Sec.~\ref{sec_alternating} to accommodate for the scaling properties. The well-understood two-lead SC-AIM allows then to validate the newly developed NRG technique in Sec.~\ref{sec_scaim_results}. Additionally, new insights into the properties of sub-gap states are obtained here. We then proceed to the gapped AIM with piece-wise constant but generally particle-hole asymmetric TDOS in Sec.~\ref{sec_gaim_results} and give a first-ever unbiased NRG solution to the problem. In Sec.~\ref{sec_gaim_compar}, a detailed comparison with existing results on gapped AIM is performed. Here, we show that NRG methods with auxiliary metallic leads provided very good results but techniques using redefined Wilson chains can only be used for qualitative assessment. In Sec.~\ref{sec_conclusions}, the main results of the paper are subsequently briefly summarized.

\section{Theory \label{sec_theory}}

\subsection{Gapped Anderson impurity model \label{sec_gapped_aim} }

For later convenience, we consider general AIM with two leads and a QD characterized by the Coulomb repulsion $U$ and the level energy $\varepsilon_d$, which in the most general case is arbitrary, but here we only consider $\varepsilon_d=-U/2$. However, we stress that the herein presented methods do not suffer from any limitations in this regard. The resulting Hamiltonian is then a sum of
\begin{eqnarray}
	H_d
	&=&
	\sum_{\sigma} 
	\varepsilon_{d}
	d^{\dagger}_{\sigma}
	d^{\vphantom{\dagger}}_{ \sigma}
	+
	U
	d^{\dagger}_{\uparrow}
	d^{\vphantom{\dagger}}_{ \uparrow}
	d^{\dagger}_{\downarrow}
	d^{\vphantom{\dagger}}_{ \downarrow},
	\label{dotH}
	\\
	H_{\alpha}  
	&=&
	\sum_{\mathbf{k}\sigma} 
	\, \varepsilon_{\mathbf{k}\alpha}
	c^{\dagger}_{\mathbf{k}\alpha \sigma}
	c^{\vphantom{\dagger}}_{\mathbf{k} \alpha\sigma},
	\label{kineticH}
	\\
	H_{T,\alpha} 
	&=&
	\sum_{\mathbf{k} \sigma} \,
	\left(V^*_{\mathbf{k}\alpha}  
	c^{\dagger}_{\mathbf{k}\alpha\sigma}
	d^{\vphantom{\dagger}}_{\sigma} 
	+ 
	V_{\mathbf{k}\alpha}  
	d^{\dagger}_{\sigma}
	c^{\vphantom{\dagger}}_{\mathbf{k}\alpha\sigma}\right),
	\label{tunnelH}
\end{eqnarray}
where $c^{\dagger}_{\mathbf{k} \alpha\sigma}$ ($c^{\vphantom{\dagger}}_{\mathbf{k}\alpha\sigma}$) creates (annihilates) an electron of spin $\sigma \in \{\uparrow \downarrow \}$, quasi-momentum $\mathbf{k}$ in lead $\alpha$ which takes values $L$ (left) or $R$ (right lead). In analogy, $d^{\dagger}_{\sigma}$ ($d^{\vphantom{\dagger}}_{\sigma}$) creates (annihilates) a dot electron of spin $\sigma$, $\varepsilon_{\mathbf{k}\alpha}$ is an unspecified dispersion relation. 

The QD hybridizes with the leads via $V_{\mathbf{k}\alpha}$, which we leave unspecified and instead prefer to extract the tunneling self-energy via
\begin{equation}
	\Sigma(\omega^+)
	=
	\sum_{\alpha \in \{L,R\}\mathbf{k}}
	\!\!\!\!\!
	V^*_{\mathbf{k}\alpha} 
	\left(
	\omega^+  - \varepsilon_{\mathbf{k}\alpha}
	\right)^{-1}
	V_{\mathbf{k}\alpha}.
	\label{eq:sigma_gapped}
\end{equation}
and then demand its imaginary part (the tunneling TDOS) to a desired form. For constant but gapped TDOS with potentially particle-hole asymmetric band, we thus require
\begin{equation}
	\Im \Sigma(\omega^+)
	=
	\begin{cases}
		0
		&
		\text{for }|\omega|<\Delta,
		\\
		\left[ 1-\mathcal{A} \, \mathrm{sgn}(\omega) \right] \Gamma_S
		&
		\text{for }|\omega|\geq\Delta,
	\end{cases}
	\label{eq:DOS_flat}
\end{equation}
where $\mathcal{A}$ governs the particle-hole asymmetry of the gapped band. In detail, for $\mathcal{A} = 0$ the TDOS remains symmetric as in Refs.~\cite{Chen-1998,Moca-2010} but for any $\mathcal{A}\neq 0$ different weights to the hole and electronic parts are ascribed.

\begin{figure*}[ht]
	\includegraphics[width=2.00\columnwidth]{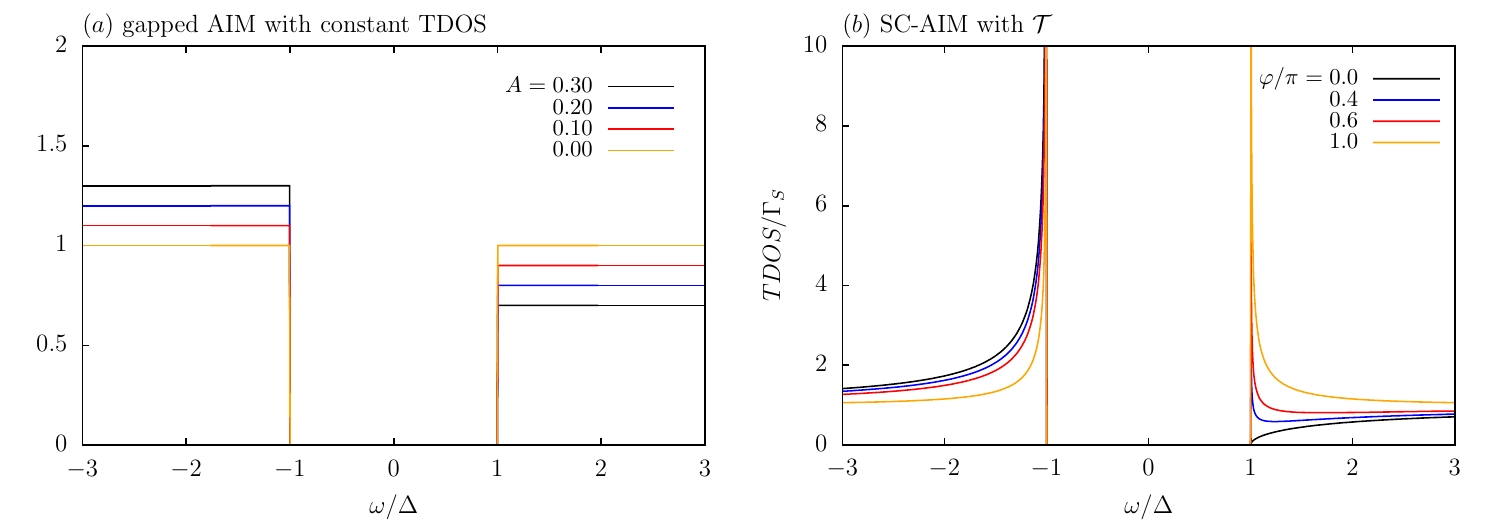}
	\caption{
		$(a)$
		TDOS functions \eqref{eq:DOS_flat} of the gapped AIM with constant TDOS at various $\mathcal{A}$. The completely particle-hole symmetric case is obtained at $\mathcal{A}=0$.
		$(b)$
		TDOS functions corresponding to SC-AIM in the scalar representation \eqref{eq:bcs_sigma_w} for various phase differences $\varphi$. The completely particle-hole symmetric case is obtained at $\varphi=\pi$
		\label{fig_tdos}}
\end{figure*}

\subsection{SC-AIM in the scalar representation \label{sec_scaim_theory}}

An essential clue toward an unified theory of gapped AIMs was provided in Ref.~\cite{Zalom-2021} by the herein employed mapping procedure of SC-AIM. To briefly summarize the approach, we define SC-AIM with one QD and two superconducting leads. Its Hamiltonian is given as a sum of the QD Hamiltonian $H_d$ according to \eqref{dotH}, two tunneling Hamiltonians $H_{T,\alpha}$ in the form of \eqref{tunnelH} and two Hamiltonians $H_{BCS,\alpha}$ describing the left ($\alpha=L$) and right ($\alpha=R$) lead via the Bardeen-Cooper-Schrieffer (BCS) theory which gives
\begin{eqnarray}
	H_{BCS}  
	&=&
	\sum_{\mathbf{k}\alpha\sigma} 
	\, \varepsilon_{\mathbf{k}}
	c^{\dagger}_{\mathbf{k} \alpha\sigma}
	c^{\vphantom{\dagger}}_{\mathbf{k} \alpha\sigma}
	\nonumber
	\\
	&-&
	\Delta
	\sum_{\mathbf{k}}
	\left(e^{i\varphi_{\alpha}} 
	c^{\dagger}_{\mathbf{k}\alpha \uparrow} 
	c^{\dagger}_{-\mathbf{k}\alpha \downarrow}
	+
	\textit{H.c.}\right),
	\label{eq:bcs}
\end{eqnarray}
where $\Delta$ is the superconducting gap and $\varphi_{\alpha}$ are superconducting order parameters in the left and right lead. For this paper, we choose symmetric hybridizations $\Gamma_L$ and $\Gamma_R$ and the gauge $\varphi_{L}=-\varphi_{R}=\varphi/2$ with $\varphi$ being the phase difference across the SC-AIM. There is however no loss of generality due to the relation between symmetric and asymmetric coupling scenario shown in Ref.~\cite{Zonda-2017}. It is now customary to employ the Nambu spinors
\begin{eqnarray}
	C^{\dagger}_{\alpha \, \mathbf{k}} 
	&=&
	\left(
	c^{\dagger}_{\alpha \, \mathbf{k} \uparrow},
	c^{\vphantom{\dagger}}_{\alpha \, -\mathbf{k} \downarrow}
	\right),
	\\
	D^{\dagger} 
	&=&
	\left(
	d^{\dagger}_{\uparrow},
	d^{\vphantom{\dagger}}_{\downarrow}
	\right).
\end{eqnarray}
Under the standard BCS assumption of $\varepsilon_{\mathbf{k}\alpha}=\varepsilon_{-\mathbf{k}\alpha}$ and with a convenient choice of real tunnel couplings $V_{\alpha\mathbf{k}}=V^*_{\alpha\mathbf{k}}=V_{\alpha-\mathbf{k}}$,  the Hamiltonians (\ref{kineticH}) and (\ref{tunnelH}), apart from possible unimportant constant energy shifts, become 
\begin{eqnarray}
	H_{\alpha} 
	&=&
	\sum_{\mathbf{k}} 
	C_{\alpha\, \mathbf{k}}^{\dagger} 
	\mathbb{E}^{\vphantom{\dagger}}_{\alpha \, \mathbf{k}}
	C_{\alpha\, \mathbf{k}}^{\vphantom{\dagger}},
	\label{NambuHalpha}
	\\
	H_{T,\alpha}
	&=&
	\sum_{\mathbf{k}}
	\left(
	D^{\dagger}
	\mathbb{V}^{\vphantom{\dagger}}_{\alpha \, \mathbf{k}} 
	C_{\alpha\mathbf{k}}^{\vphantom{\dagger}}
	+
	C_{\alpha\mathbf{k}}^{\dagger} 
	\mathbb{V}^{\vphantom{\dagger}}_{\alpha \, \mathbf{k}} 
	D^{\vphantom{\dagger}}
	\right)
	\label{NambuHtunel}
\end{eqnarray}
with
\begin{eqnarray}
	\mathbb{E}_{\alpha\mathbf{k}}  
	& = &
	-\Delta_\alpha C_{\alpha} \sigma_x
	+\Delta_\alpha S_{\alpha} \sigma_y
	+ \varepsilon_{\mathbf{k}\alpha} \sigma_z,
	\label{Ealpha}
	\\
	\mathbb{V}_{\alpha\mathbf{k}}
	& = &
	V_{\alpha\mathbf{k}}  \, \sigma_z,
	\label{Valpha}
\end{eqnarray}
where  $\sigma_i$, $i\in \{x,y,x\}$, are the Pauli matrices 
and $C_{\alpha} \equiv \cos{\varphi_\alpha}$, $S_{\alpha} \equiv \sin{\varphi_\alpha}$. The blackboard bold typeface distinguishes matrices from scalars. The corresponding TDOS is then given as
\begin{equation}
	\mathbb{\Sigma}^D(\omega^+)
	=
	\sum_{\alpha \in \{L,R\}\mathbf{k}}
	\!\!\!\!\!\mathbb{V}_{\alpha\mathbf{k}}
	\left(
	\omega^+ \mathbb{1} - \mathbb{E}_{\alpha\mathbf{k}}
	\right)^{-1}
	\mathbb{V}_{\alpha\mathbf{k}}
\end{equation}
which yields
\begin{equation}
	\mathbb{\Sigma}^D(\omega^+)
	=
	\Gamma_S
	\left[	
	\omega \mathbb{1} 
	+ \Delta \cos\left(\frac{\varphi}{2}\right) \sigma_x
	\right]
	F(\omega^+)
	\label{eq:bcs_sigma_d}
\end{equation}
with
\begin{equation}
	\begin{split}
		F(\omega^+)
		&\equiv
		\frac{1}{\pi}
		\int_{-B}^{B}
		\frac{d\varepsilon}{\omega^2 - \Delta^2 - \varepsilon^2 + i\eta\, \mathrm{sgn}(\omega)}\\
		&=\frac{1}{\pi\sqrt{(\omega+ i\eta)^2 - \Delta^2 }}\ln\frac{\sqrt{(\omega+ i\eta)^2 - \Delta^2}+B}{\sqrt{(\omega+ i\eta)^2 - \Delta^2}-B}.
	\end{split}
\end{equation}

Taking the $\eta \rightarrow 0$ limit, we arrive at 
\begin{equation}
	F(\omega^+)
	=
	\begin{cases}
		-\frac{2}{\pi\sqrt{\Delta^2-\omega^2}} \arctan
		\left(\frac{B}{\sqrt{\Delta^2-\omega^2}}\right),
		&
		\text{for }|\omega|<\Delta
		\\
		-\frac{i\, \mathrm{sgn}(\omega)}{\sqrt{\omega^2-\Delta^2}}
		+
		\frac{\ln \left( \frac{B+\sqrt{\omega^2-\Delta^2}}{B-\sqrt{\omega^2-\Delta^2}}\right)}{\pi\sqrt{\omega^2-\Delta^2}},
		&
		\text{for }\Delta<|\omega|<B.
	\end{cases}
\end{equation}
The resulting $\mathbb{\Sigma}^D(\omega^+)$ has thus a non-zero imaginary part only for $|\Delta|<\omega$ while all effects of the finite-sized band appear in its real part. However, once the limit $B\rightarrow\infty$ is taken the real 
part out of the gap vanishes too. 

To bring the SC-AIM model into a form of a general gapped AIM we perform the unitary transformation $\mathbb{T}$
\begin{eqnarray}
	\mathbb{T} D
	&\equiv&
	W,
	\qquad  \, \, \, \, \, \, \, \,
	D^{\dagger} \mathbb{T}^{\dagger}
	\equiv
	W^{\dagger},
\end{eqnarray}
where
\begin{equation}
	\mathbb{T}
	=
	\frac{1}{\sqrt{2}}
	\left(
	\sigma_x - \sigma_z
	\right)
	\qquad
	\text{and}
	\qquad
	W ^{\dagger}
	=
	\left(
		w^{\vphantom{\dagger}}_{\uparrow}
		\, \,
		w^{\dagger}_{\downarrow}
	\right).
	\label{eq_transf}
\end{equation}
The resulting TDOS in basis of $W$ spinors is then
\begin{equation}
	\mathbb{\Sigma}^W(\omega^+)
	=
	\Gamma_S
	\left[	
	\omega 
	-
	\Delta \cos\left(\frac{\varphi}{2}\right) 
	\right]
	F(\omega^+).
	\label{eq:bcs_sigma_w}
\end{equation}
Taking the imaginary part of \eqref{eq:bcs_sigma_w} yields the TDOS for SC-AIM in the scalar representation with typical BCS divergences appearing at the edges while $\varphi$ modulates the resulting particle-hole asymmetry in the basis of the $w$ fields as shown in Fig.~\ref{fig_tdos}.

Finally, at half-filling the $H_d$ Hamiltonian is form invariant against the transformation $\mathbb{T}$ and reads
\begin{eqnarray}
	H_d
	&=&
	\sum_{\sigma} 
	\varepsilon_{d}
	w^{\dagger}_{\sigma}
	w^{\vphantom{\dagger}}_{ \sigma}
	+
	U
	w^{\dagger}_{\uparrow}
	w^{\vphantom{\dagger}}_{ \uparrow}
	w^{\dagger}_{\downarrow}
	w^{\vphantom{\dagger}}_{ \downarrow}.
\end{eqnarray}

\section{NRG approach for gapped TDOS functions \label{sec_nrg}}

The here presented novel NRG approach generalizes the seminal work~\cite{Bulla-1994} to the gapped TDOS. We thus start in Sec.~\ref{sec_gapped_tdos} by generalizing the proof given in Ref.~\cite{Bulla-1994} to the gapped TDOS scenario. While the mathematical proof is relative straightforward, it yields only limited specifications on the discretization procedure itself. Nevertheless, studying discretized self-energy as in Sec.~\ref{sec_loggap_disc}, allows for a preliminary justification to the here employed log-gap discretization. The resulting Wilson chains incorporate two separate scales in regimes of $T \gg \Delta$ and $T \ll \Delta$ respectively which require significant modifications to the kept/discarded state scheme of the NRG diagonalizations as discussed in Sec.~\ref{sec_alternating}. We stress that $z$-averaging techniques can be potentially incorporated into the presented approach to improve especially the spectral functions. However, in the present paper we focus on subgap spectroscopy and the validation of the developed NRG approach. We therefore set $z=1$ throughout this work to streamline the presentation.

\subsection{Gapped TDOS in approach of R. Bulla \label{sec_gapped_tdos} }

Let us assume that an impurity problem with gapped TDOS is present and its tunneling self-energy is given as
\begin{eqnarray}
	\Sigma(z)
	=
	\int_{I_{\Delta}}
	\sum_{\mu}
	d\omega
	V^2(\omega) \rho(\omega) G_{\mu}(\omega)
	\label{eq:app_sigma_gapl}
\end{eqnarray}
with $G_{\mu}(\omega)$ being the Green's function of the lead with a gapped DOS $\rho(\omega)$, while $I_{\Delta}$ denotes an integration interval running from $-B$ to $B$ with a spectral gap of width $2\Delta$ that is centered at the Fermi energy of $0$. An example of such a self-energy is provided, for example, by Eq.~\eqref{eq:bcs_sigma_w}. Our aim is now to construct a corresponding one-channel Hamiltonian 
\begin{eqnarray}
	H
	-
	H_d
	&=&
	\int_{I_{\Delta}}
	\sum_{\mu}
	d\varepsilon
	\,
	g(\varepsilon)
	\,
	a^{\dagger}_{\mu}
	a_{\mu}
	\nonumber
	\\
	&+&
	\int_{I_{\Delta}}
	\sum_{\mu}
	d\varepsilon
	\,
	h(\varepsilon)
	\,
	d^{\dagger}_{\mu}
	a_{\mu}	+H.c.
	\label{eq:app_H-Hd_gap}
\end{eqnarray}
with $g(\varepsilon)$ and $h(\varepsilon)$ to be specified in such a way that the tunneling self-energy defined by \eqref{eq:app_H-Hd_gap} is identical to \eqref{eq:app_sigma_gapl}. We thus first change the integration variable from $\varepsilon$ to $g$ and solve for the tunneling self-energy of \eqref{eq:app_H-Hd_gap}. We obtain
\begin{eqnarray}
	\Sigma(z)
	=
	\int_{I_{\Delta}}
	\sum_{\mu}
	dg
	\,
	h^2[\varepsilon(g)] \frac{d\varepsilon}{dg}
	G_{\mu}(g).
	\label{eq:app_sigma_gapr}
\end{eqnarray}
Comparing \eqref{eq:app_sigma_gapl} and \eqref{eq:app_sigma_gapr} implies the following relation between $h(\varepsilon)$ and $g(\varepsilon)$:
\begin{eqnarray}
	V^2(x) \rho(x)
	=
	h^2[\varepsilon(x)] \frac{d\varepsilon}{dx},
	\label{eq:app_condition_gap}
\end{eqnarray}
which holds for every $x \in I_{\Delta}$ and has exactly the form known from Ralf Bulla's seminal work~\cite{Bulla-1994}. This is of crucial importance for gapped TDOS problems as the discretization step can be performed without the introduction of any approximations. We may simply choose, for example, a piecewise constant $h(\varepsilon)$ on intervals $I^{\pm}_n$, where $+$ denotes intervals in positive and $-$ in negative frequency domain. To reconstruct the TDOS we can then transfer all required details onto $g(\varepsilon)$ via condition \eqref{eq:app_condition_gap}. 

However, Eq.~\eqref{eq:app_sigma_gapr} gives us no prescription on how to choose the discretization intervals $I^{\pm}_n$, it just requires them to reside within the interval $I_{\Delta}$. This leaves us with two main options so $n$ might be either a finite or infinite set of integer numbers. Since the tridiagonalization equations are of the same form as for the ordinary NRG, as applied for example to AIM with metallic bath, a finite number of discretization intervals would leave us with a finite Wilson chain, which will inevitably have some smallest built-in energy scale. Moreover, the concept of approaching a low-temperature fixed point for Wilsonian RG iterations will make no sense for at all.

To avoid such problems, one simply takes the iterative structure of the tridiagonalization equations into account by defining the initial discretization intervals $I_0^{\pm}$ at the band edges $\pm B$. As $n$ is increased the intervals patch the remainder of $I_{\Delta}$ and get smaller towards the gap edges at $\pm\Delta$. In formal agreement with Ref.~\cite{Bulla-1994}, we then define the following quantities
\begin{eqnarray}
	\xi_n^{+/-}
	=
	\frac{\int_{I^{+/-}_n} dx \, x \Gamma(x)}{\int_{I^{+/-}_n} dx \Gamma(x)},
	\label{eq_xi_coeff}
	\\
	\left( \gamma_n^{+/-} \right)^2
	=
	\int_{I^{+/-}_n} dx \, \Gamma(x),
	\label{eq_gamma_coeff}
\end{eqnarray}
where $\Gamma(x)$ is the imaginary part of \eqref{eq:app_sigma_gapl}. These enter then subsequently the tridiagonalization equations (28)-(31) given in Ref.~\cite{Bulla-Rev-2008}. In detail, each interval $I_n^{+/-}$ gives rise to operators $a^{\dagger}_{n,\sigma,p}$ for $I_n^{+}$ and $b^{\dagger}_{n,\sigma,p}$ for $I_n^{-}$. These are connected to a set of orthonormal functions indexed by $\pm$, $n$ and also $p$, where the last index takes all integer values. Inserting these into the corresponding Hamiltonians and taking piecewise constant approximation of the hybridization term in the energy representation allows only $p=0$ components, so we can completely drop the $p$ indices and obtain the corresponding discretized Hamiltonian as
\begin{eqnarray}
H 
=
H_{imp}
&+&
\sum_{n,\sigma}
\left(
\xi_n^+
a^{\dagger}_{n,\sigma}
a^{\vphantom{\dagger}}_{n,\sigma}
+
\xi_n^-
b^{\dagger}_{n,\sigma}
b^{\vphantom{\dagger}}_{n,\sigma}
\right)
\nonumber
\\
&+&
\sum_n
\left(
\gamma_n^+
d^{\dagger}
a^{\vphantom{\dagger}}_{n,\sigma}
+
\gamma_n^-
d^{\dagger}
b^{\vphantom{\dagger}}_{n,\sigma}
\right)
\nonumber
\\
&+&
\sum_n
\left(
\gamma_n^+
a^{\dagger}_{n,\sigma}
d^{\vphantom{\dagger}}
+
\gamma_n^-
b^{\dagger}_{n,\sigma}
d^{\vphantom{\dagger}}
\right).
\label{eq_discretized}
\end{eqnarray}

\subsection{Log-gap discretization for gapped AIMs \label{sec_loggap_disc} }

\begin{figure*}[ht]
	\includegraphics[width=2.10\columnwidth]{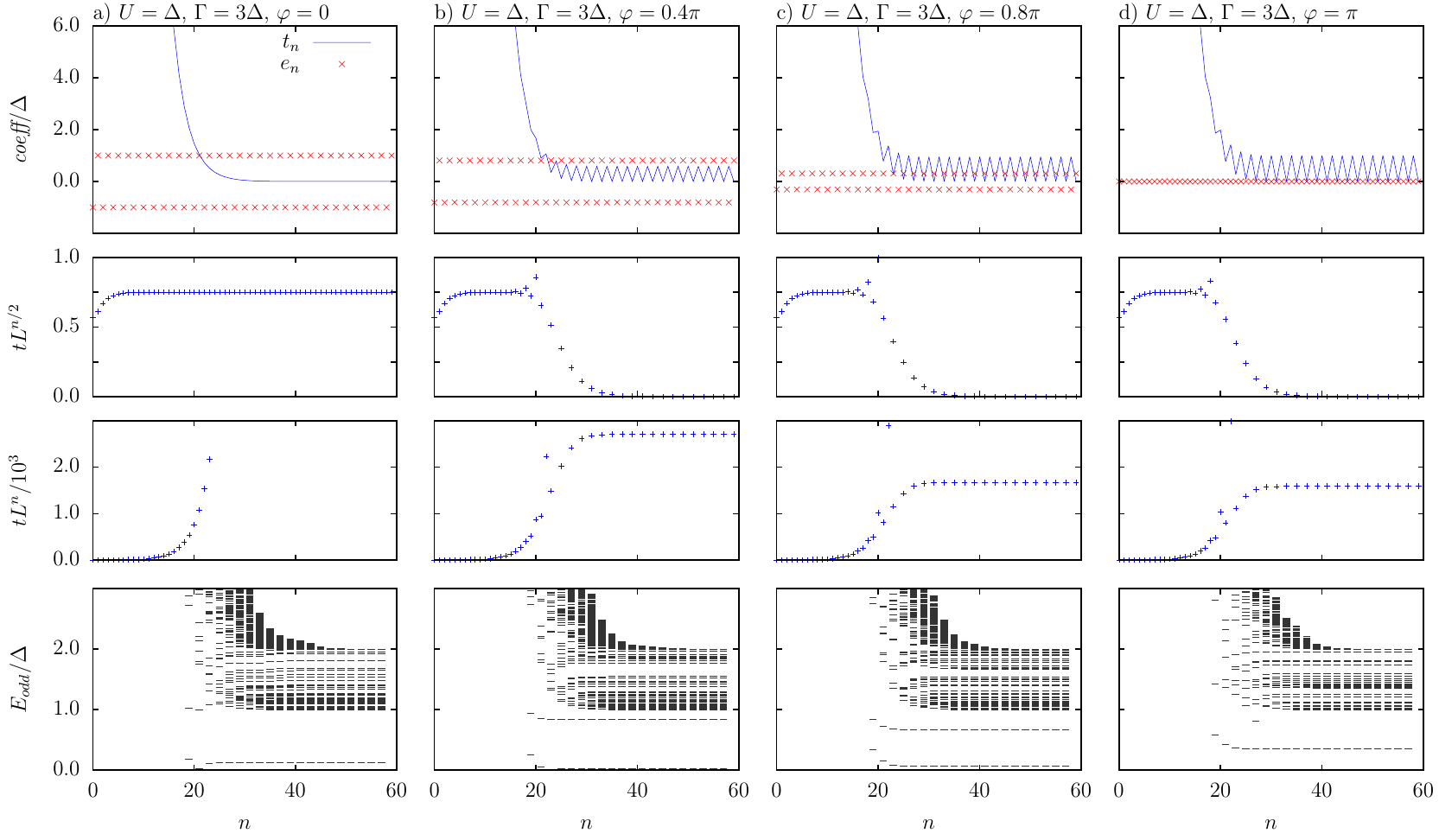}
	\caption{
		$(a)$-$(d)$
		Hoppings $t_n$ (solid blue line), on-site energies $\epsilon_n$ (red crosses) and the corresponding energy eigenvalues (gray lines) of the Wilson chain calculated via the log-gap approach for SC-AIM with $U=3\Delta$ and $\Gamma=\Delta$ at several phase differences $\varphi$.
		First row shows unrescaled values of Wilson chain parameters. The onsite energies $\epsilon_n$ are alternating in the sign, but their absolute value is approximately constant. However, hoppings $t_n$ follow two successive scalings.
		The first one with $t_n \sim \Lambda^{-n/2}$ is shown in the second raw and applies to high temperatures ($T \gg \Delta$). Upon decreasing the phase bias toward $\varphi=0$ the region of high temperature scaling increases until it dominates the entire temperature range for $\varphi=0$.
		The third raw shows then the low temperature ($T \ll \Delta$) scaling of hoppings obeying $t_n \sim \Lambda^{-n}$, which sets in after the short crossover region of approximately ten NRG iterations.
		The last raw shows eigenenergies obtained using the log-gap NRG approach with diagonalization scheme according to Sec.~\ref{sec_alternating}. Correct energy scale separation is observed.
		\label{fig_2_wilson}}
\end{figure*}

To guide us in the choice of the precise form of the discretization intervals $I^{\pm}_n$ we solve now for the discretized tunneling self-energy $\Sigma_{disc}(z)$ of \eqref{eq_discretized}. It reads
\begin{eqnarray}
	\Sigma_{disc}(z)
	=
	\sum_n
	\frac{-\left( \gamma_n^+ \right)^2}{\xi_n^+ - z}
	+
	\frac{\left( \gamma_n^- \right)^2}{-\xi_n^- + z}.
	\label{eq_disc_sigma}
\end{eqnarray}
The two branch cuts of the continuous $\Sigma(z)$ that follow the real axis and terminate at $\pm \Delta$ have thus been replaced by simple isolated poles at positions $\xi_n^+>0$ and $\xi_n^-<0$. The coefficients \eqref{eq_xi_coeff} have thus a straightforward meaning of the positions of the poles, while \eqref{eq_gamma_coeff} defines the corresponding weights in the discretized TDOS. 

Analogous results are present also in other commonly used approaches to NRG. For example, in AIM with metallic reservoir, the continuous self-energy with branch cut over the entire real axis gives rise under the ordinary logarithmic discretization to simple poles that accumulate logarithmically around the Fermi energy. From the perspective of gapped systems, they accumulate exactly where the two branch cuts over positive and negative part of the real axis merge as $\Delta = 0$. Similar pattern is also observed for the superconducting bath, which is the only system with a spectral gap that is currently rigorously treated by NRG and thus closest to the problem advanced here. Once again, using standard two-channel NRG techniques for SC-AIM the discretized tunneling self-energy $\Sigma_{disc}(z)$ is obtained in the form of \eqref{eq_disc_sigma} with simple poles emerging along the two branch cuts. Significantly, they are known to accumulate logarithmically at the terminal points $z=\pm \Delta$ of the branch cuts.

Our aim now is therefore to construct a discretization that is not only in line with the requirements of the previous section, but reproduces additionally the logarithmic accumulation of simple poles in the corresponding tunneling self-energy $\Sigma_{disc}(z)$ at $z=\pm \Delta$ where the branch cuts of the continuous self-energy terminate. To this end we define the intervals $I^{\pm}_n$ via their end points $x^{\pm}_n$ as
\begin{equation}
x_n^{\pm}
=   
\pm
\left[
\Delta + (1-\Delta) \Lambda^{-n}
\right],
\label{eq_log_gap_disc}
\end{equation}
where $\Lambda$ is the usual discretization parameter with intervals $I^{\pm}_n$ obviously aggregating at the gap edges in a logarithmic way (bandwidth set as $2B \equiv 2$). Consequently, we refer to \eqref{eq_log_gap_disc} as the log-gap discretization. We stress that it was already considered in Ref.~\cite{Liu-2016}, however, the scaling properties of the resulting Wilson chains have been completely missed and the employed standard kept/discarded scheme must have resulted in ill converging solutions at low temperatures, which the authors did not report in the manuscript. As shown in the next section, further modifications to the NRG approach are required to resolve this problem. 

Let us, now calculate the pole structure of the discretized self-energy. First, we consider the gapped Anderson model with TDOS \eqref{eq:DOS_flat}. We note that the discretized self-energy becomes of form \eqref{eq_disc_sigma} with poles $\xi_n^{+/-}$ and weights $\gamma_n^{+/-}$ obeying
\begin{eqnarray}
	\xi_n^{+/-}
	&=&
	\pm
	\left[
	\Delta + \frac{(1-\Delta) (1 + \Lambda)}{2} \Lambda^{-1 - n}
	\right],
	\label{eq_xi_coeff_flat}
	\\	
	\left(
	\gamma_n^{+/-}
	\right)^2
	&=&
	(1 \pm \mathcal{A}) (1-\Delta) (\Lambda-1) \Lambda^{-1 - n}.
	\label{eq_gamma_coeff_flat}
\end{eqnarray}
Setting $\Delta=\mathcal{A}=0$, one recovers the standard case of the AIM with constant metallic TDOS, where poles approach the gap edges in a logarithmic fashion. The same holds true also for $\Delta\neq0$ at arbitrary $\mathcal{A}$, only these time the poles are logarithmically approaching the gap edges as in the standard NRG approach to SC-AIM. The log-gap discretization is thus increasingly sensitive to the states which get closer to the Fermi energy, albeit due to the presence of the gap, it can never be reached. Notably, poles \eqref{eq_xi_coeff_flat} are placed symmetrically around the Fermi energy regardless of the values of $\mathcal{A}$ and/or $\Delta$. Consequently, the particle-hole asymmetry is completely encoded only by the weights $\gamma_n^{+/-}$. 

Since the constant gapped TDOS is structureless at the gap edges, let us also corroborate the SC-AIM case in its scalar representation as shown in Fig.~\ref{fig_tdos}. The BCS-like divergences on the gap edges and a complicated particle-hole asymmetry of \eqref{eq:bcs_sigma_w} leads to the following expression
\begin{eqnarray}
\xi_n^{+/-}
&=&
\pm
\left[
\Delta + f^{\pm}(\varphi,\Delta,\Lambda) \Lambda^{-1 - n}
\right]
\end{eqnarray}
where the prefactors $f^{\pm}(\varphi,\Delta,\Lambda)$ are functions of $\Delta$ and $\Lambda$ and are generally not particle-hole symmetric, i. e. $f^{+}(\varphi,\Delta,\Lambda) \neq f^{-}(\varphi,\Delta,\Lambda)$, unless $\varphi=\pi$. Consequently, the overall particle-hole asymmetry of \eqref{eq:bcs_sigma_w} is distributed over poles and weights of the discretized self-energy, causing an effect that needs to be addressed later. Nevertheless, the logarithmic behavior toward the terminal points of the branch cuts is still preserved. Nevertheless, compared to the standard SC-AIM solution the poles are not distributed in a particle-hole asymmetric fashion, which is an effect that needs to be considered later as a potential source of differences.

Next, we will feed the poles and weights into the tridiagonalization equations to obtain the corresponding Wilson chains. Due to the corroborated differences to the only known standard NRG solution we choose the SC-AIM case for these demonstrations. We select $\Gamma=\Delta$, $U=3\Delta$, $\varepsilon_d=-U/2$ and vary $\varphi$. The resulting parameters of Wilson chains for selected values of $\varphi$ are then shown in the first row of Figs.~\ref{fig_2_wilson}$(a)$-$(d)$, while second and third row demonstrate high- and low-temperature scalings of the hoppings respectively.

As shown in the first row of Figs.~\ref{fig_2_wilson}$(a)$-$(d)$, at large $n$ even hoppings become approximately $\Delta \sin\varphi$, while odd hoppings become exponentially suppressed. To decipher the various scalings present in the system, let us look at the second and the third row of Figs.~\ref{fig_2_wilson}$(a)$-$(d)$. Obviously, there is always an initial stage where hoppings scale in an usual $\Lambda^{-n/2}$ way, which however survives down to $n\rightarrow \infty$ only for $\varphi=0$. Otherwise, the scaling is interrupted by a crossover region and only odd hoppings scale exponentially with $\Lambda^{-n}$ law.

The behavior of the on-site energies $\epsilon_n$ on the other hand is more straightforward as they vary between the approximate values of $\pm \Delta cos(\varphi)$. The alternating pattern holds approximately for all model parameters when log-gap NRG approach is applied with only particle-hole symmetric TDOS leading to the vanishing of all on-site energies. 

While the presented cases were calculated for SC-AIM, we stress that the alternating feature of the on-site energies and the two-scaled nature of hoppings are observed also for the gapped AIM model with constant TDOS function. However, here the second stage scaling involves odd hoppings to follow the $\Lambda^{-n/2}$ law. So clearly, the presence of the additional BCS-like divergencies in the TDOS of the SC-AIM case modifies the power law of the odd hoppings in the low-temperature regime.

\section{Iterative NRG diagonalization of two-scaled Wilson chains \label{sec_alternating}}

\begin{figure}[t]
	\includegraphics[width=1.00\columnwidth]{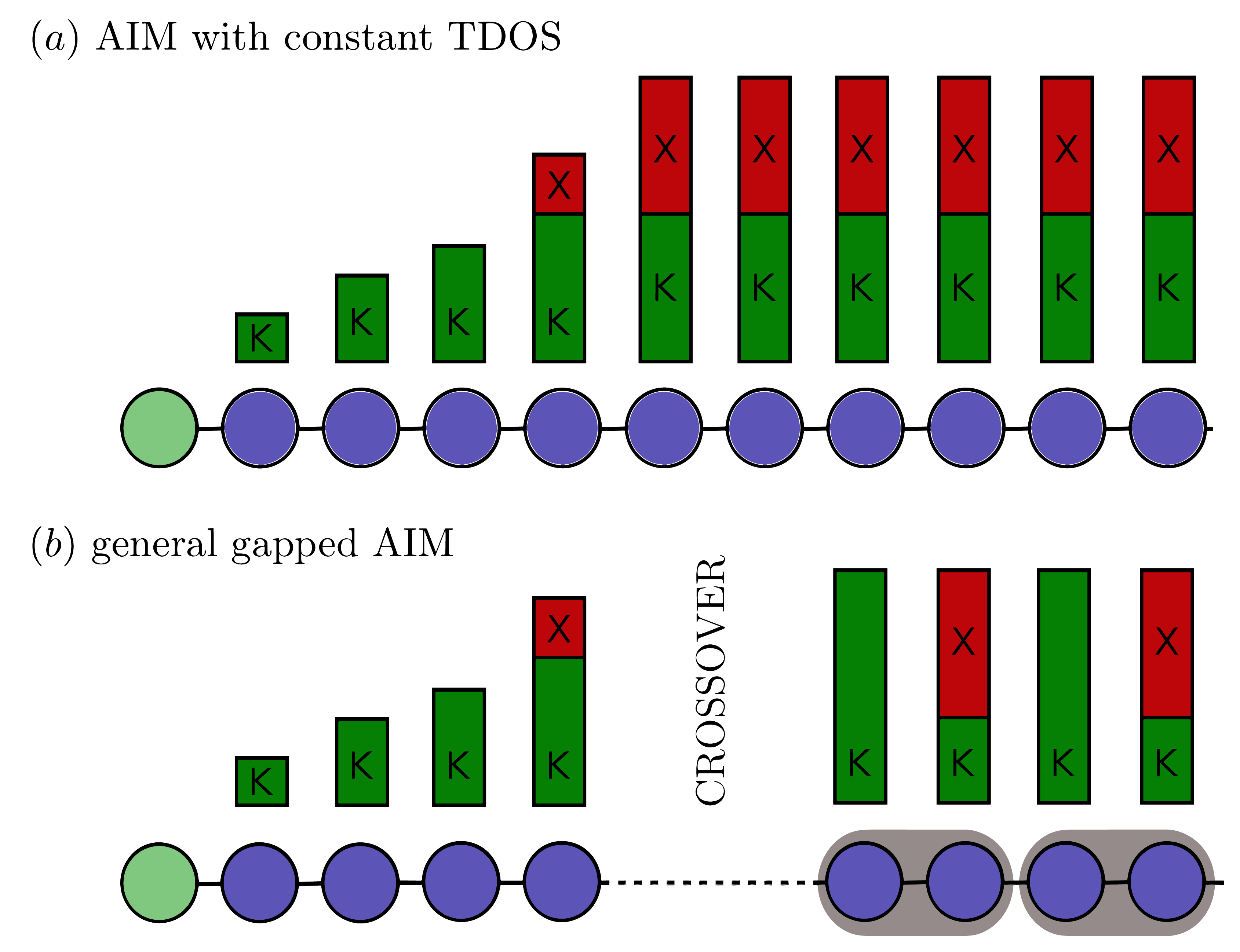}
	\caption{
		$(a)$
		An illustration of the outcome of the ordinary strategy to keep (green, letter $K$) or discard (red, letter $X$) states (or multiplets) during iterative NRG diagonalization of AIM with metallic TDOS. After few initial iterations a maximum number of kept states is quickly reached.
		$(b)$ 
		The Wilson chain of a general gapped AIM incorporates an initial scaling of odd and even hoppings as $t_n \sim \Lambda^{-n/2}$ which undergoes a smooth crossover into the second stage scaling, where odd hoppings scale as $\Lambda^{-n}$ and the even ones reach a constant value. To ensure energy scale separation, it is thus necessary to keep all states at odd NRG iterations. Such an alternating kept/discarded scheme reinterprets thus the Wilson chain into a tight-binding Hamiltonian of non-interacting serial double QDs.
		\label{fig_3_kd}}
\end{figure} 

The observed two-scaled nature of the hoppings is an essential feature which needs to be adequately considered during the iterative NRG diagonalization procedure. To begin with, let us however first discuss the special case of $\varphi=0$ which exhibits only the ordinary $\Lambda^{-n/2}$ scaling for all hoppings and was first described in the seminal papers on NRG \cite{Wilson-1975,Krishna-1980a,Krishna-1980b}. Iterative addition of the sites of the Wilson chains leads to an exponential increase of the total Hilbert space which is dealt with by a truncation strategy as illustrated in Fig.~\ref{fig_3_kd}$(a)$. Thus, already after few NRG iterations high energy states (multiplets) are being discarded systematically with a more or less constant number kept until the algorithm is stopped. 

Such a diagonalization procedure works only in conjunction with the $\Lambda^{-n/2}$ scaling of all hoppings, which ensures energy scale separation (high and low eigenenergy states do not mix as the chain grows). While originally established for impurities in metallic baths, the same scaling and consequently energy separation emerges also for one-lead SC-AIM problems \cite{Satori-1992}. Quite surprisingly and without an apparent justification, it was also established for the two-lead SC-AIM, where the addition of complex phases mixes both leads together \cite{Hecht-2008}. 

In our case, such mixing is also observed once $\varphi \neq 0$ and leads to the second stage of scaling, where even hoppings become of the order $\Delta\cos\varphi$ and odd hopping scale as $\Lambda^{-n}$. In this second stage, we can thus naturally reinterpret the Wilson chain as being composed of serial double QDs interconnected by the exponentially decreasing hopping. If the Hilbert space is never truncated after the odd site is added, splittings of eigenenergies by factor proportional to $\Lambda^{-n}$ have to occur at even iterations and energy scale separation can be exploited. 

As demonstrated graphically in Fig.~\ref{fig_3_kd}$(b)$, such a kept/discarded strategy is perfectly tailored for the low energy scaling, but it would be difficult to be turned on only after one emerges from the initial scaling. Fortunately, when applied also to the high temperature portion of the Wilson chain, the novel kept/discarded strategy just keeps superfluous high energy states (multiplets) every odd iteration. At even iterations, they give then rise to states which are immediately discarded, so in principle, the novel kept/discarded strategy is just inefficient in the $\Lambda^{-n/2}$ scaling sector but does not break the energy scale separation as demonstrated in the fourth row of Figs.~\ref{fig_2_wilson}$(a)$ by the $\varphi=0$ case where second stage scaling is not present at all. Here, eigenenergies at odd NRG iterations $E_{odd}$ (no rescaling applied), clearly converge as $n$ is increased up to $60$. Modifications to the kept/discarded strategy in the low-temperature sector are thus not necessary and can be potentially reverted to speed up the numerics, which however requires careful investigations. Such refinements are left for future developments.

For al other values of $\varphi$ the second stage scaling occurs after a crossover region, but as shown in the fourth row of Figs.~\ref{fig_2_wilson}$(b)$-$(d)$ the here developed kept/discarded strategy indeed ensures energy scale separation. We also stress, that at every NRG iteration new low energy states resurface around the gap edge (not including the subgap states) and get exponentially close to the region right above the gap edge as $n$ is increased. Their appearance does not however alter the other higher lying eigenvalues, which are already converged. This is, of course, a feature well-known also from the standard NRG solution of SC-AIM as discussed in the Appendix of Ref.~\cite{Satori-1992}.

The new strategy of keeping or discarding states comes with the price of increased CPU and memory requirements over the traditional scheme of Fig.~\ref{fig_3_kd}$(a)$ since the intermediate Hilbert space used for matrix diagonalizations grows by factor of up to $16$, which thus resembles the ordinary two-channel NRG calculations. However, in the scalar basis used here, charge and pseudo-spin quantum numbers label the states (multiplets) and the increased symmetry is crucial for speeding up the diagonalizations when the log-gap NRG approach is employed. Consequently, the here proposed NRG scheme for SC-AIM remains feasible at any modern desktop also when $\varepsilon \neq -U/2$ and even in the presence of non-zero magnetic field, while standard two-channel NRG solution reaches borders of its practical feasibility.

Apparently, the same kept/discarded strategy is also required for the AIM with constant but gapped TDOS, albeit the odd hoppings of the second stage scaling follow the $\Lambda^{-n/2}$ law. Still the even hoppings alternate between positive and negative values. Consequently, the low-temperature portion of the Wilson chain has once again the character of a tight-binding chain composed of serial double QDs connected by exponentially decreasing hoppings. The application of the log-gap NRG with the modified kept/discarded strategy then finally also confirms the working energy scale separation.

While the kept/discarded scheme of Fig.~\ref{fig_3_kd}$(b)$ is crucial for performing unbiased Wilsonian RG calculations for the here discussed models, we would like to briefly explore the outcome when an ordinary kept/discarded scheme according to Fig.~\ref{fig_3_kd}$(a)$ is employed even for Wilson chains of the two-scaled nature. The energy scale separation is then, of course, not ensured and the eigenspectrum of energies become corrupted for $\varphi \neq 0$ or $\varphi \neq \pi$ as demonstrated in Appendix~\ref{sec_app_fast}. Nevertheless, problems occur initially only in its high energy part and the sub-gap spectrum remains largely intact, so intermediate NRG iterations can be used to determine sub-gap properties with only few percent of relative error when compared to the rigorous Wilsonian approach developed here. We will call this approach the approximate log-gap NRG, albeit we stress that it is actually a misnomer as energy scale separation is broken. It should thus strictly be used only for fast scannings and, at least, a portion of the data should always be validated against rigorous methods. Additionally, supragap spectral functions or thermodynamic properties depend upon high-energy states and will thus be susceptible to much larger systematic errors. In the main part of this paper, we will therefore use only the here developed rigorous log-gap NRG and will briefly discuss the veracity and possibilities of its fast but approximate version in Appendix~\ref{sec_app_fast}.

\section{Results} 

\subsection{SC-AIM \label{sec_scaim_results} }

\begin{figure}[t]
	\includegraphics[width=1.00\columnwidth]{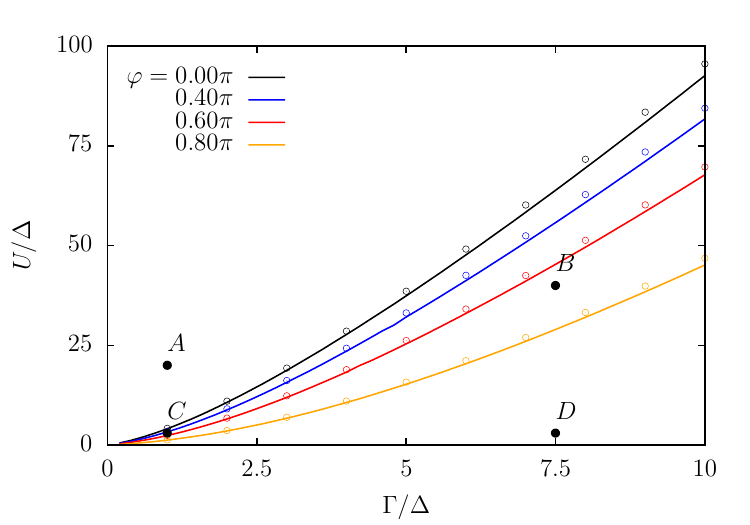}
	\caption{
		$(a)$
		The phase transition lines of SC-AIM calculated via the log-gap NRG (solid lines) and the standard two-channel NRG approach (points) at four different values of $\varphi$. Doublet (singlet) phase is realized above (below) the lines. The difference between both methods is $\approx 3\%$ and steady for all shown data points. The comparison for the phase bias dependent position of ABS states is undertaken for the four selected cases $A$, $B$, $C$ and $D$ in Fig.~\ref{fig_abs}.
		\label{fig_phasebound_bcs}}
\end{figure} 

\begin{figure*}[ht]
	\includegraphics[width=2.00\columnwidth]{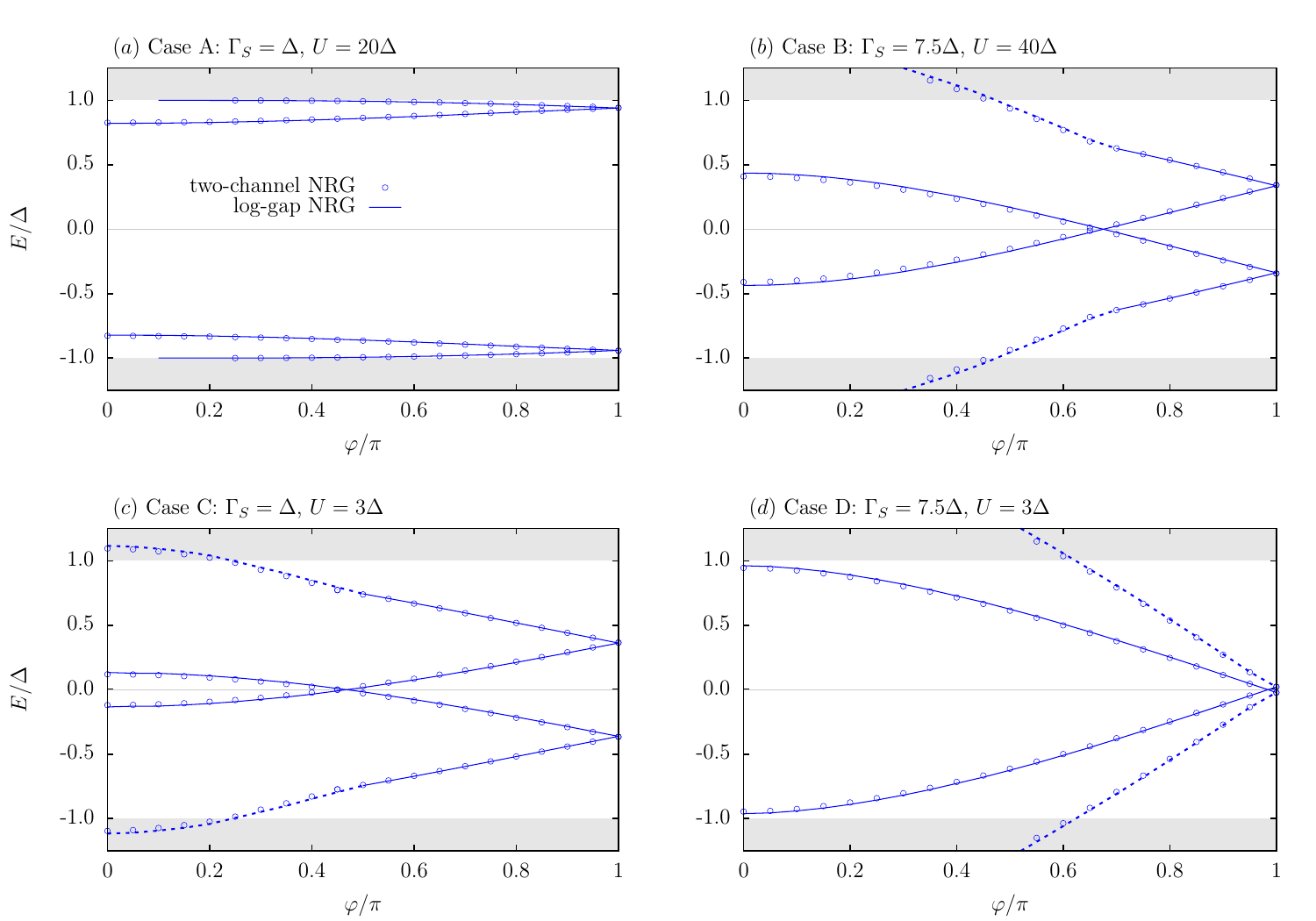}
	\caption{
		$(a)$-$(d)$
		$\varphi$-dependent behavior of sub-gap excitations for four selected cases $A$, $B$, $C$ and $D$ of SC-AIM with parameters according to Fig.~\ref{fig_phasebound_bcs}. 
		Solid lines represent the ABS states, while dashed lines show the two-particle excitations calculated by the log-gap NRG approach at $\Lambda=2$.
		The reference data was calculated using the two-channel NRG with $\Lambda=4$ (open circles). 		
		The agreement is overall very good, but a systematic tendency of increased differences occurs as $\varphi$ tends towards $0$. The escape of sub-gap states into the continuum portion of the spectrum was rigorously verified in cases $B$, $C$ and $D$ via the log-gap NRG approach and only subsequently the corresponding states from the two-channel NRG could be identified.
		\label{fig_abs}}
\end{figure*} 

In the previous sections, a rigorous Wilsonian NRG scheme has been developed to tackle arbitrary problems with QD immersed in a gapped TDOS. Before solving the gapped Anderson model, as an example, we first validate the log-gap NRG against the well-established two-channel standard solution of SC-AIM by using its scalar representation derived in Ref.~\cite{Zalom-2021}. We will also demonstrate its effectivity in regard to the CPU time by obtaining the full phase diagram and we will exploit the quantum numbers for tracking of in-gap excitations as they leave into the continuum, which was currently speculated upon in Ref.~\cite{Pokorny-2023}. 

Starting by obtaining the complete phase diagram for SC-AIM with QD at half-filling in Fig.~\ref{fig_phasebound_bcs}, we notice that the resulting singlet-doublet QPT lines divide the $\Gamma/\Delta$-$U\Delta$ plane into the upper doublet GS portion and the lower singlet GS part. Four selected values of $\varphi$ are shown with $\varphi=\pi$ leading to the exclusive doublet GS phase for all parameters as long as $\varepsilon_d=-U/2$. Due to the employed symmetries in the log-gap NRG solution of SC-AIM, the calculations come only at the fraction of the CPU time required by the ordinary NRG and are feasible on any modern desktop. We have thus pre-calculated the data in Fig.~\ref{fig_phasebound_bcs} via the log-gap approach and used these as an ansatz for the standard NRG calculations.

The resulting parity transition lines obtained using the log-gap NRG approach with $\Lambda=2$ (solid lines) are then compared to the standard two-channel NRG calculations at $\Lambda=4$ (points). While only a small and steady difference of $\approx 3\%$ between both methods is observed, we select four cases $A$, $B$, $C$ and $D$ at various ratios of $U/\Gamma$ for a detailed study. The resulting sub-gap spectroscopy including one- and two-particle excitations is then presented in Fig.~\ref{fig_abs}. Once again the log-gap NRG approach with $\Lambda=2$ (solid and dashed lines for one- and two-particle excitations respectively) and the standard two-channel NRG with $\Lambda=4$ (open circles) are generally in a very good agreement, but the difference systematically grows as $\varphi$ is decreased. This hints toward an increasing importance of the asymmetric pole structure of the self-energy due to the log-gap discretization, while the effect of different $\Lambda$ appears less significant.

In case $A$, the ratio $U/\Gamma = 20$ drives the system into the doublet GS phase for all possible values of $\varphi$. An opposite scenario at small ratio of $U/\Gamma$, case $D$, is then conversely marked by singlet GS phase dominating almost the entire phase evolution. The doublet GS phase is however always present, albeit only in a small region around $\varphi \approx \pi$ due to the phenomenon of the doublet chimney as explained in Ref.~\cite{Zitko-arxiv}. Cases $B$ and $C$ have on the other hand moderate ratios $U/\Gamma$, which results in a comparably similar phase regions of singlet and doublet GS phase. 

Additionally, the newly developed NRG technique assigns charge and pseudo-spin numbers to the NRG eigenstates and allows thus direct tracking of the in-gap excitations as they cross into the continuum region in cases $B$, $C$ and $D$. Only case $A$ is an exception, since the continuous part of the spectrum and the state corresponding to the excitation of interest  share similar quantum numbers. The directly observed values have then been used to cross-identify the most likely candidates from the standard two-channel NRG calculations. Our observations are in good accord with indirect observations in Ref.~\cite{Pokorny-2023}.

\begin{figure}[t]
	\includegraphics[width=1.00\columnwidth]{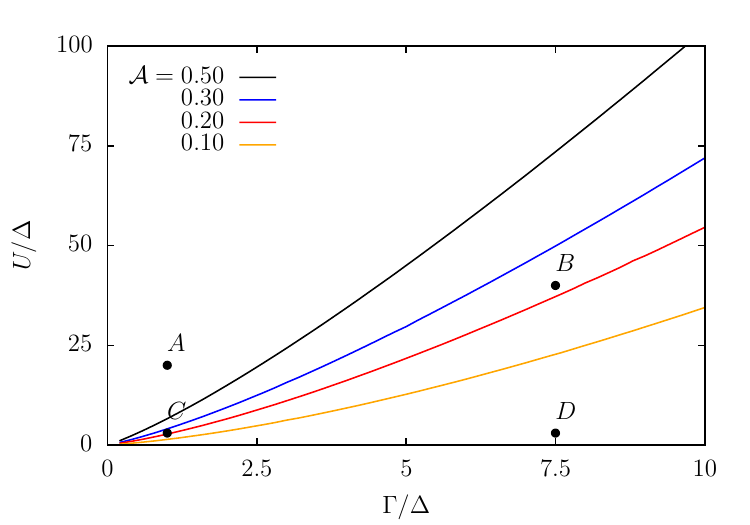}
	\caption{
		Phase boundaries between the singlet and doublet GS for four selected values of particle-hole asymmetry parameter $\mathcal{A}$ of the gapped AIM with constant TDOS. The case of $\mathcal{A}=0$ results in purely doublet GS and is thus not shown. Nevertheless, the phase diagram clearly indicates that even a proportionally small amount of particle-hole asymmetry is enough to cause a transition into a singlet phase at large hybridizations $\Gamma$. The phase diagram also clearly indicates relation to the SC-AIM shown in Fig.~\ref{fig_phasebound_bcs}.
		\label{fig_phasebound_gaim}}
\end{figure}

\subsection{Gapped AIM with constant TDOS \label{sec_gaim_results} }

Successful benchmarking of the log-gap NRG approach against the standard NRG results allows us to proceed with the gapped AIM problem with constant TDOS and give its full unbiased NRG solution. In this section, we present the observed physical phenomena and emphasize their origin in the shared RG-equivalence class with the SC-AIM. Detailed comparisons to previous attempts on gapped AIM with auxiliary metallic leads or methods with redefined Wilson chains are presented separately in Sec.~\ref{sec_gaim_compar}. Here, we only state that a very convincing agreement with NRG solutions with auxiliary metallic lead is observed, while the results of Refs.~\cite{Chen-1998,Moca-2010,Pinto-2015} are proven as highly approximate.

For now, let us start with the phase diagram of the problem as shown in Fig.~\ref{fig_phasebound_gaim}. Using the log-gap discretization, the resulting Wilson chains for gapped AIMs are of one channel nature and can be diagonalized in a standard iterative approach of NRG. The resulting GS parity can be read off directly, unlike in NRG methods with auxiliary metallic leads. The hypothesis about the GS and in-gap parities of Refs.~\cite{Zalom-2022} are nevertheless finally directly confirmed. Consequently, GS transition lines divide the $U/\Delta$-$\Gamma/\Delta$ plane into an upper doublet and lower singlet portion. Overall, the similarity to the phase diagram of SC-AIM in Fig.~\ref{fig_phasebound_bcs} is exemplary. 

\begin{figure*}[ht]
	\includegraphics[width=2.00\columnwidth]{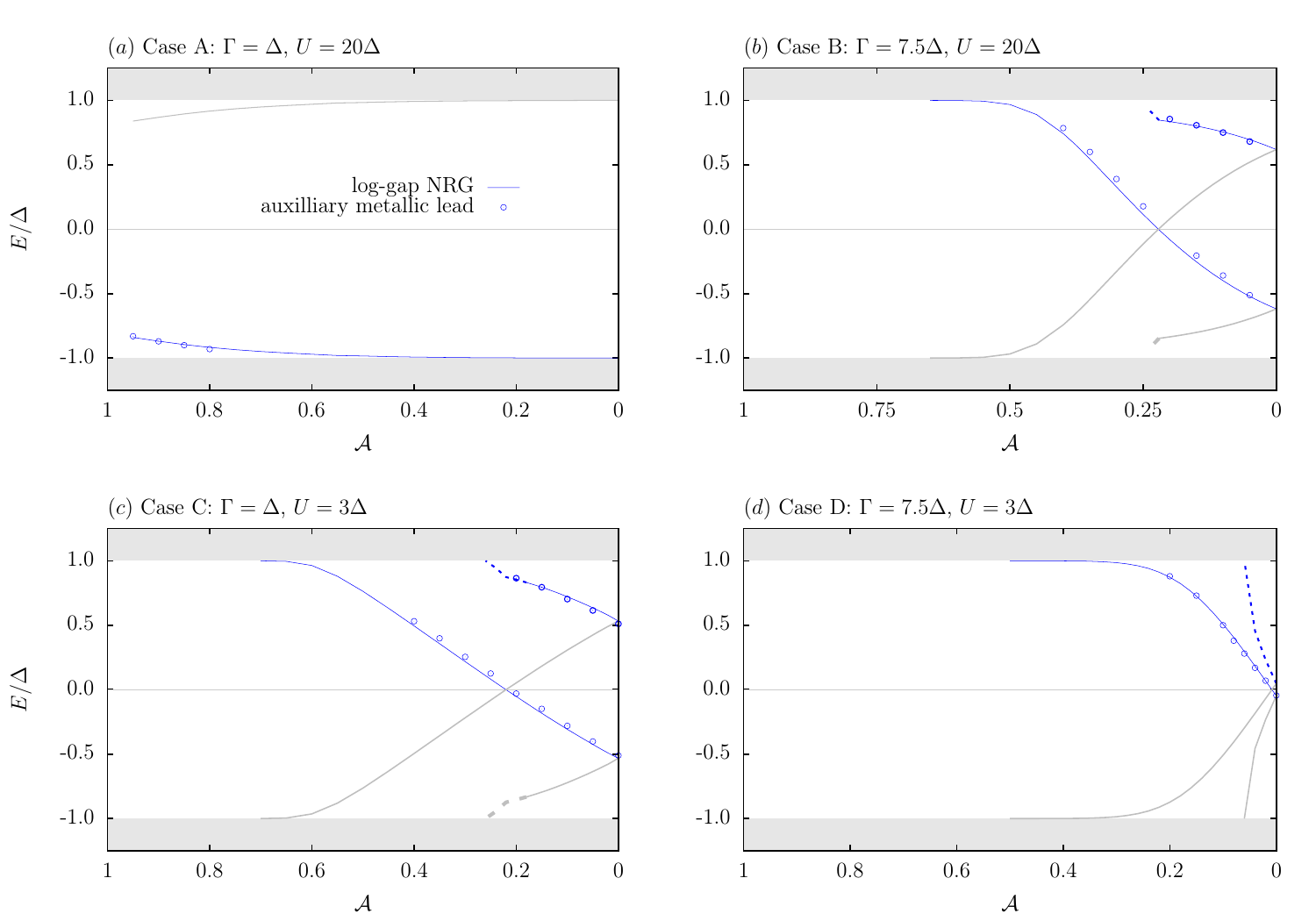}
	\caption{
		$(a)$-$(d)$
		$\mathcal{A}$-dependent behavior of sub-gap excitations for four selected cases $A$, $B$, $C$ and $D$ of the gapped AIM with constant TDOS according to Fig.~\ref{fig_phasebound_gaim} calculated with the log-gap NRG at $\Lambda=2$. Solid and dashed lines represent one- and two-particle excitations, respectively. The solid thin gray lines show the result for the model with $\mathcal{A} \rightarrow -\mathcal{A}$.
		Results are compared with the NRG solution for the model augmented with auxiliary metallic lead (open circles) as performed in Ref.~\cite{Zalom-2022}. Due to restrictions of the later method only one-particle excitations could be determined.
		Using the one- and two-particle excitations, no in-gap states have been found to escape into the continuum, which is the main difference to the SC-AIM case shown in Fig.~\ref{fig_abs}.
		\label{fig_bound_gaim}
	}
\end{figure*}

The particle-hole asymmetry $\mathcal{A}$ of the gapped band plays thus an analogous role as $\varphi$ in SC-AIM in its scalar representation. In detail, $\mathcal{A}=0$ and $\varphi=\pi$ correspond to each other and represent the particle-hole symmetric choices for the corresponding models. Consequently, at $\mathcal{A}=0$ the whole $U/\Delta$-$\Gamma/\Delta$ plane supports only the doublet GS, giving rise to the phenomenon of doublet chimney as discussed in Ref.~\cite{Zitko-arxiv} specifically for SC-AIM. Nevertheless, inducing even a small particle-hole asymmetry $\mathcal{A}$ causes a QPT into the singlet GS at large hybridizations $\Gamma/\Delta$. The larger the particle-hole asymmetry $\mathcal{A}$ becomes, the more extended the singlet phase space becomes. All of these findings point then to a deeper connection between SC-AIM and other gapped AIMs as hypothesized already in Ref.~\cite{Zalom-2022}. In the present approach, the connection is already evident after the discretization step is performed, as the two-scaled Wilson chain emerges for both models.

To further deepen these findings, we select four cases $A$, $B$, $C$ and $D$ with exactly the same $U/\Delta$ and $\Gamma/\Delta$ as in the SC-AIM case. We then calculate the evolution of the in-gap excitations with respect to the particle-hole asymmetry parameter $\mathcal{A}$ as shown in Fig.~\ref{fig_bound_gaim} and use solid (dashed) lines to track the one-particle (two-particle) excitations. We stress that unlike in SC-AIM, in-gap excitations do not come in pairs except of $\mathcal{A}=0$, but this is a trivial consequence of gapped AIM being directly defined in the basis of  (generally) particle-hole asymmetric fermions. Anyways, to make the correspondence to the SC-AIM problem more obvious we also add excitations which appear for the model upon $\mathcal{A} \rightarrow -\mathcal{A}$, but we stress that they never appear simultaneously.

In detail, the large Coulomb interaction in case $A$ expels the singlet in-gap states to the edge of the subgap region and ensures doublet GS for all values of $\mathcal{A}$. At $\mathcal{A}=0$, the underlying symmetry causes both singlet excited states to overlap, which appears then as two symmetrically placed one-particle excitations in the sub-gap spectral function. At $\mathcal{A}\neq 0$, one of these singlets is pinned to the gap edge and starts moving closer to the Fermi energy only as $\mathcal{A} \approx 0.5$. The other singlet crosses then into the continuum part of the spectrum. The only difference to the SC-AIM case lies thus in a different curvature of the observed trajectories, which is of quantitative nature and clearly depends on the shape of the TDOS. The case $D$ shows then in analogy an contrasting outcome to case $A$. Due to the large hybridization $\Gamma$, already a very small particle-hole asymmetry of $\mathcal{A} \approx 0.01$ suffices to induce a doublet-singlet QPT. Consequently, singlet phase dominates the in-gap states. 

The outcome in cases $B$ and $C$ is once again in close analogy to the SC-AIM results with the initial doublet GS being present up to moderate values of $\mathcal{A}\approx 0.25$. The first excited state at $\mathcal{A}=0$ is of doublet nature and appears in hole as well as particle part of the spectral function. Increasing then $\mathcal{A}$ causes both in-gap peaks of the spectral function to move in the same direction towards the gap edge. Together with their symmetric counterparts (thin gray lines), which physically appear together only in SC-AIM, we observe a typical pattern. Once again, up to the missing symmetrization both models behave qualitatively in an analogous way. These findings are thus consistent with the hypothesis of Ref.~\cite{Zalom-2022}, where both models are stipulated to belong into the same RG universality class.

\subsection{Comparison with older results for gapped AIM \label{sec_gaim_compar} }

In the literature, there are in principle two alternative NRG-based proposals for solving gapped AIM problems. One relies on Wilson chains that follow from corresponding AIM cases with $\Delta=0$ \cite{Chen-1998, Moca-2010,Pinto-2015}. In the other approach, standard NRG methods are used but the gapped system is augmented by a weakly coupled metallic lead as used in Refs.~\cite{Diniz-2020,Zalom-2021,Zalom-2021r,Zalom-2022}.

The main idea for the former approach lies in first constructing a Wilson from the $\Delta=0$ case via the standard logarithmic discretization. Its length is set to $M$ so that the last hopping coefficient equals approximately $\Delta$. The resulting tight-binding chain is then solved by an usual NRG iterative diagonalization and the results are considered to represent the gapped case. In other words, closing the gap just means solving an ordinary Anderson model at ever-growing Wilson chain.

Apparently several problems do arise with such a construction. Taking any finite $\Delta$, we note that the logarithmic discretization (as any other running up to the Fermi energy) inevitably crosses into the gap region at some given $N$ and the Wilson chain should terminate here. But the characteristic energy scale of the smallest hopping is then $\Lambda^{-N/2}$ which is approximately $\sqrt\Delta$ and thus much larger then the desired energy scale, or in other words, the previously chosen length $M$ is actually equal to $2N+1$. The problem with such an approach lies therefore not in truly the finiteness of the Wilson chain but in its generalization beyond the length $N$. In the end, it is thus questionable why a Wilson chain of the ungapped problem with one particular length $2N+1$ (which is mathematically not even permissible) should represent the gapped scenario. 

On the other hand, the other commonly used method uses only standard NRG techniques and modifies only the model definition by adding a weakly coupled metallic lead fill the gap region with a small but non-zero TDOS. One then calculates a series of spectral functions at decreasing couplings $\Gamma_M$ of the added metallic lead and analyzes the sub-gap peaks which are present as broadened analogs of the sharp in-gap peaks. 

The $\Gamma_M\rightarrow 0$ limit is then deduced from such a series, which is is time consuming and inaccurate as a non-zero numeric threshold on smallest possible $\Gamma_M$ exists. Moreover, there is no guarantee for the $\Gamma_M=0$ case to adiabatically connect to this series with supporting evidence being only circumstantial for the SC-AIM problem \cite{Zalom-2021,Zalom-2021r}. Additionally, this indirect method is not able to give indisputable conclusions about the quantum numbers of the GSs and in-gap states in question. Taking together, both existing methods have to be treated with caution and do not represent a rigorous Wilsonian RG approach to the fully gapped problem. 

To proceed, we will thus subject the two approaches and the here developed method to a scrutiny. To this end, we have selected the three cases from Fig.~2 in Ref.~\cite{Moca-2010} which correspond to the gapped Anderson impurity model with constant TDOS with parameters according to Table.~\ref{tab_pascu}. To avoid any inconsistencies in the application of the approach used by the authors, the positions of the in-gap peaks have been read off as indicated in Table.~\ref{tab_pascu}. Second, using the method of auxiliary metallic lead we have recalculated the spectral functions and determined the peak positions under the consideration. The resulting values are also shown in Table.~\ref{tab_pascu}. Note that closing the gap at fixed $U$ and $\Gamma$ causes the in gap states to approach the Fermi energy, where a Kondo peak develops due to the auxiliary metallic lead. The accurate reading of the in-gap peak position becomes therefore increasingly complicated. Nevertheless, such a problem arises mainly for the case of $\Delta=7.05 \times 10^{-6}B$, while all in-gap positions determined in Ref.~\cite{Moca-2010} are systematically larger than with the auxiliary lead by a factor of up to $2$.

Finally, all three cases can be solved using the here presented log-gap NRG. As already predicted in Fig.~\ref{fig_bound_gaim}, in-gap peaks are located well inside the gap region, which is sharply separated from the continuous part. The extracted peak positions are then presented in Table.~\ref{tab_pascu}. Clearly, their correspondence to the method of auxiliary metallic lead is good, while result from Ref.~\cite{Moca-2010} are off by a factor of up to $2$. The still obvious differences between log-gap NRG and the method using an auxiliary metallic lead grow as $\Delta$ is decreased. Here, the method of auxiliary metallic lead becomes increasingly difficult as at the smallest numerically possible value of the coupling $\Gamma_M$ the in-gap peak was still moving toward the gap edge and thus possibly towards the position determined by the log-gap NRG.

\begin{figure}[ht]
	\includegraphics[width=1.00\columnwidth]{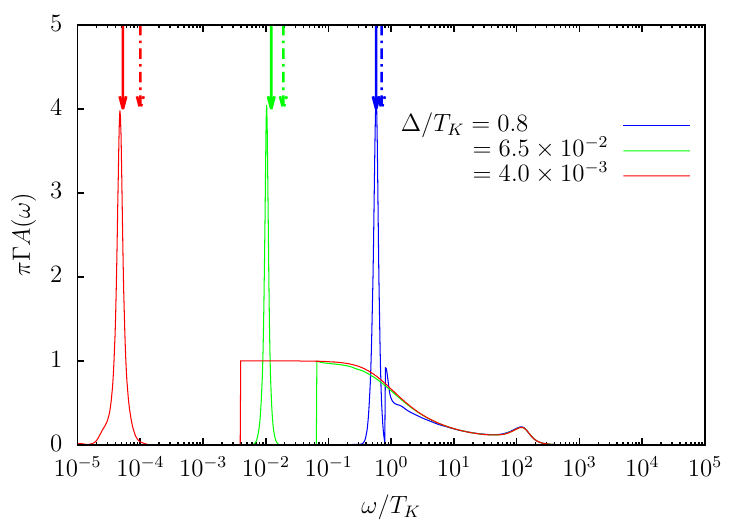}
	\caption{
		Solid lines show spectral functions calculated with the auxiliary metallic lead according to Refs.~\cite{Diniz-2020,Zalom-2021,Zalom-2021r,Zalom-2022}. Solid arrow shows the position of the in-gap peaks calculated using the log-gap approach while dashed arrows are placed positions corresponding to Fig.~3 of Ref.~\cite{Moca-2010}. Note the overall good equivalence between the log-gap and auxiliary metallic lead method, while values using the approach of Refs.~\cite{Chen-1998,Moca-2010} are clearly misaligned with the other two results. Parameters for the AIM with constant but gapped TDOS are $U=4B$, $\Gamma=0.4B$, $\mathcal{A}=0$ and $\Delta/B$ varying according to the legend.
		\label{fig_pascu}}
\end{figure}

Moreover, in Fig.~2 of Ref.~\cite{Moca-2010} one observes the continuous part of the spectral functions to spill over the gap edges into the gap region. We thus conclude that the method used in Ref.~\cite{Moca-2010} is highly approximate and inconsistent even on its own. Contrary, in Ref.~\cite{Zalom-2022} in-gap peaks at $\mathcal{A}=0$ have always appeared inside of a well-defined gap which is also confirmed after recalculating the three cases of Ref.~\cite{Moca-2010} as demonstrated in Fig.~\ref{fig_pascu}. The same holds true for the results from the log-gap NRG. 

We also stress that the mutual correspondence of the log-gap NRG and the method of auxiliary metallic lead is not confined to the three selected $\mathcal{A}=0$ cases as evident from the result presented in Fig.~\ref{fig_bound_gaim} with a discussion about the comparison with other methods being postponed up to this moment. Clearly, there are differences perceptible even with a bare eye, but one has to bear in mind that the auxiliary metallic lead method is inherently plagued by several systematic and numeric problems. First, a limit of $\Gamma_M \rightarrow 0$ is hard to reach as a smallest value of numerically allowed $\Gamma_M$ still broadens the in-gap peaks and reading off the in-gap positions is then associated with errors. Additionally, in-gap peaks close to the gap edges are overlapping with the gap edges. Consequently, most of the $\mathcal{A}$ point for the case $A$ have not even been accessible with the method of auxiliary metallic lead. 

Nevertheless, taking together, with the exception of the case $A$ in Fig.~\ref{fig_bound_gaim}, all data points are in good agreement with the here presented log-gap NRG approach and we can thus safely conclude, that the generalized Wilson chains according to Refs.~\cite{Chen-1998, Moca-2010,Pinto-2015} are not well representing the actual AIM with constant but gapped TDOS but the remaining two approaches do so with the log-gap NRG being, of course, superior not only as the method to solve the truly gapped scenario, but also in terms of practical usability and CPU requirements.

	\begin{table}[]
	\begin{tabular}{ p{2.0cm} | p{1.9cm} | p{1.9cm} | p{1.9cm} }
		$\Delta/B$   &  $7.05 \times 10^{-6}$  & $1.15 \times 10^{-4}$  &  $1.41 \times 10^{-3}$   \\ \hline
		P. Moca in \cite{Moca-2010}:  $|E/B|$   &  $1.76 \times 10^{-7}$  & $3.35 \times 10^{-5}$  &  $1.23 \times 10^{-3}$   \\ \hline
		auxiliary: $|E/B|$  &  $8.29 \times 10^{-8}$  & $1.82 \times 10^{-5}$  &  $1.02 \times 10^{-3}$                 \\ \hline
		log-gap:  $|E/B|$   &  $9.29 \times 10^{-8}$  & $2.15 \times 10^{-5}$  &  $1.06 \times 10^{-3}$ 
	\end{tabular}
	\caption{Position of in-gap peaks for AIM with constant but gapped TDOS wit $U=4\Delta$, $\Gamma=0.4\Delta$, $\mathcal{A}=0$ and $\Delta/B$ varying accordingly. Values taken from Ref.~\cite{Moca-2010} have been compared to the log-gap method developed in the present paper and the method with auxiliary metallic lead adapted from Refs.~\cite{Diniz-2020,Zalom-2021,Zalom-2021r,Zalom-2022}. \label{tab_pascu} }
	\end{table}
	
\section{Conclusions \label{sec_conclusions} }

We have reported on a novel NRG scheme that lifts the so-far existing limits on solving impurity problems immersed in scalar but gapped TDOS via unbiased Wilsonian RG techniques. The herein developed method is referred to as the log-gap NRG approach, since it replaces the standard logarithmic discretization that runs up to the Fermi energy with an analogous procedure that accumulates the discretization points logarithmically at the gap edges. While the subsequent tridiagonalization is standard, the resulting Wilson chains contain generally two scalings which require specific diagonalization procedure, where truncation of the Hilbert space is performed only at every even NRG iteration (initial iteration enumerated as $0$-th).

Both, the log-gap discretization and the alternating kept/discarded scheme for truncation of the Hilbert space during the iterative solution of the Wilson chain, are inherently connected. The discretization procedure is motivated by the analysis of the poles in the discretized tunneling self-energy in Sec.~\ref{sec_loggap_disc} and gives rise to two-scaled behavior of the Wilson chain, which requires different truncation strategy of the resulting Hilbert space so that the energy scale separation is not broken. The two-scaled nature of SC-AIM and the gapped AIM hints towards the underlying RG-universality class.

Using the novel NRG technique, first a thorough benchmarking was performed using SC-AIM in its scalar representation as obtained in Ref.~\cite{Zalom-2021}. The results by the log-gap NRG approach compare in high numeric accord to standard NRG methods, but we stress that the log-gap NRG approach is superior in terms of CPU requirements due to the available symmetries. This holds true also when $\varepsilon_d\neq -U/2$ or non-zero magnetic field on the QD are considered. The generalization to encompass such cases is straightforward and of interest for current experiments where the log-gap NRG method can provide much faster scanning tool in the parameter space of SC-AIM (especially when its approximate version from Appendix~\ref{sec_app_fast} is employed).
 
The log-gap NRG approach was then finally used in the first ever unbiased Wilsonian RG solution of the AIM with constant but gapped TDOS. The resulting physical quantities are thus valid without any restrictions at arbitrary temperatures corresponding to the given length of the chain. The most general conclusion from these calculations is that the model behaves qualitatively as SC-AIM as they both belong to the same RG-equivalence class. In particular, one observes always a doublet GS at completely particle-hole symmetric scenario with $\mathcal{A}=0$ and $\varepsilon_d =-U/2$, which corresponds to the $\varphi=\pi$ scenario of SC-AIM. Nevertheless, dependent on the values of $\Gamma$ and $U$ it may eventually undergo a QPT into a singlet GS as $\mathcal{A}$ is increased, which corresponds to decreasing $\varphi$ in SC-AIM. A complete phase diagram in half-filled QD scenario is shown in Fig.~\ref{fig_phasebound_gaim}, which is highly similar to the corresponding phase diagram of SC-AIM calculated in Fig.~\ref{fig_phasebound_bcs}. 

In Sec.~\ref{sec_gaim_compar}, an elaborate comparison has been performed which shows that the existing results from Refs.~\cite{Chen-1998,Moca-2010,Pinto-2015} are not only in conflict with the here developed log-gap NRG method, but crucially they show also significant numeric discrepancy when compared to the standard NRG method for a system augmented with an auxiliary metallic lead as demonstrated in fig.~\ref{fig_pascu}. Contrary, the herein developed log-gap NRG matches very well. Additional reasons for the approach of Refs.~\cite{Chen-1998,Moca-2010,Pinto-2015} to be considered with caution is of theoretic origin and was laid down also in Sec.~\ref{sec_gaim_compar}.

We stress that the method is general and not restricted to SC-AIM or AIM with constant but gapped TDOS. Future generalizations and developments can incorporate the $z$-averaging techniques to solve for the spectral function, while addition of out-of-half-filling scenario or the presence of magnetic field is almost trivial. Additionally, an asymmetrically placed gap around the Fermi energy, as can appear in realistic TDOS functions, can also easily be incorporated. Another natural avenue for important applications of the here presented method represents the problem of metal-insulator transition, where the herein proposed NRG scheme might be implemented as internal impurity solver in DMFT calculations when the gap is finally open in the system. However, precaution is always required as the presence of the second stage scaling is a necessary prerequisite for meaningful application of RG techniques. Moreover, some additional problems might arise when the gap edge is not sharply defined as already noted in \cite{Zalom-2022}. For realistic scenarios, fusion with the adaptive mesh approach according to \cite{Zitko-adapt} is also necessary.  

As a final consequence of the presented approach, we would like to accentuate the missing rigorous RG theory of effective model for SC-AIM. While zero-band width or atomic limit theory are nowadays routinely employed for qualitative analysis of SC-AIM and related experiments \cite{Meng-2009,Grove-Rasmussen-2018,Meden-2019,Zonda-2023}, we are still not able to fully understand how and why they emerge from the full problem. The two-scaled nature of the Wilson chains, as deciphered here, could however allow to build a rigorous analytic RG theory of fixed points in the spirit of seminal works by K. G. Wilson \cite{Wilson-1975, Krishna-1980a,Krishna-1980b}.

Taking together, one of the main limitations of the NRG approach has been successfully eliminated, so an unbiased Wilsonian RG approach for gapped systems is now able to resolve all energy scales at the same footing. Further generalization and development is however necessary to go beyond the basic models implemented here.

\begin{acknowledgements}
	We acknowledge discussions with T. Novotn\'{y}, M. \v{Z}onda, K. Wrze\'sniewski and R. \v{Z}itko. This work was supported by Grant  No. 23-05263K of the Czech Science Foundation.
\end{acknowledgements}

\appendix

\section{Approximate version of the log-gap NRG approach  \label{sec_app_fast} }

\begin{figure*}[ht]
	\includegraphics[width=2.00\columnwidth]{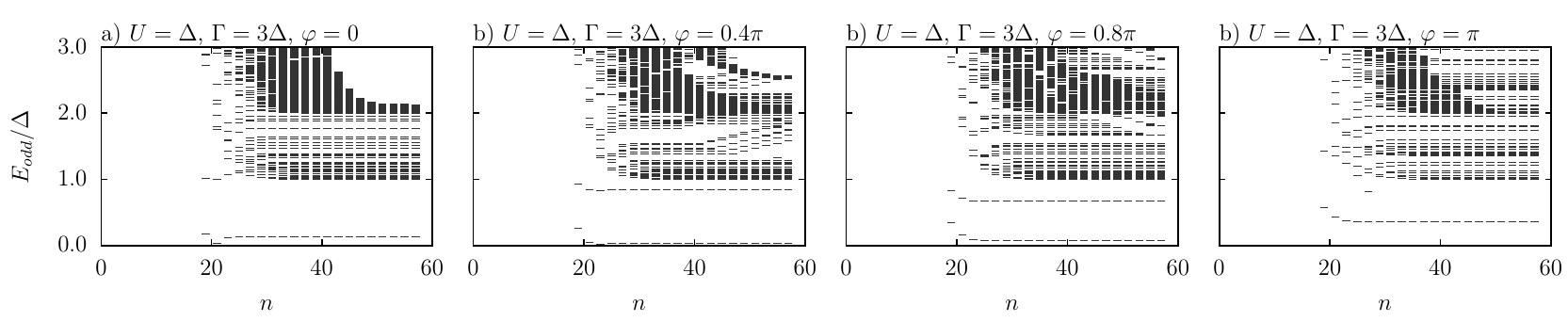}
	\caption{
		$(a)$-$(d)$
		$\varphi$-dependent behavior of NRG eigenspectra when the Wilson chains of SC-AIM with corresponding parameters obtained by the log-gap discretization are subject to the usual iterative diagonalization in which states are discarded at every iteration. Obviously, $\varphi=0$ and $\varphi=\pi$ cases do not suffer from corruption, but for all other choices states in the middle of the supragap spectrum experience unphysical flow which is only a numeric artifact.
		\label{fig_app_spectra}}
\end{figure*} 

\begin{figure*}[ht]
	\includegraphics[width=2.00\columnwidth]{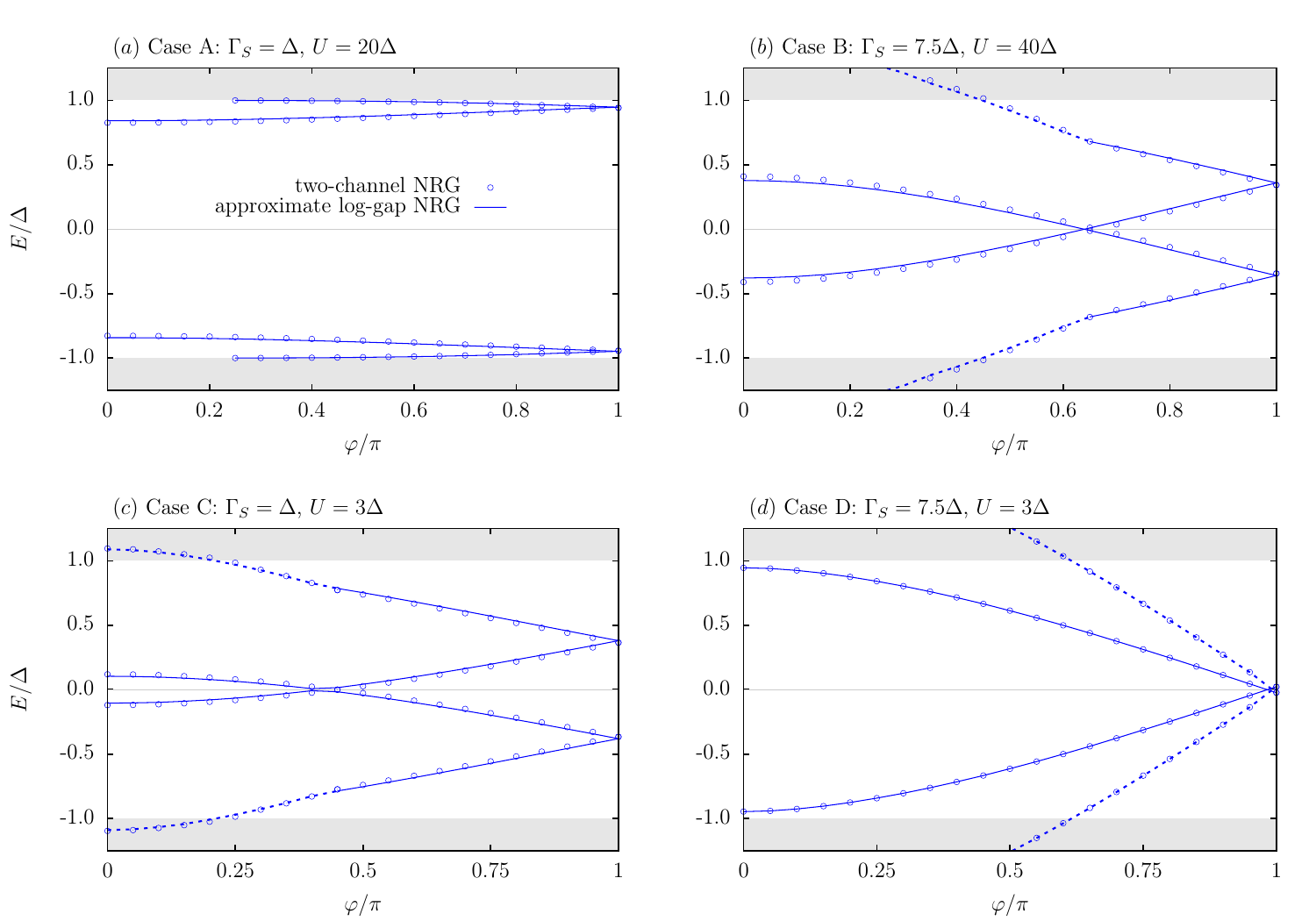}
	\caption{
		$(a)$-$(d)$
		$\varphi$-dependent behavior of in-gap states for four selected cases $A$, $B$, $C$ and $D$ of SC-AIM according to Fig.~\ref{fig_phasebound_bcs} obtained using log-gap discretization with usual Wilson chain diagonalization procedure where states are (eventually) discarded after every iteration. While the eigenspectra are generally corrupted as shown in Fig.~\ref{fig_app_spectra}, the subgap portion is altered only marginally and can be used for quick scans for QPTs. We stress, however, that thermodynamic quantities or spectral functions would suffer from systematic errors and this approximate approach should be discouraged in such instances.
		\label{fig_app_fast_abs}}
\end{figure*} 

In the main text only the results of a rigorous Wilsonian RG approach are used. However, as briefly mentioned in Sec.~\ref{sec_alternating}, an approximate and a much faster calculation (in terms of the used CPU time) can be used for either preliminary scanning of large parametric spaces or for obtaining subgap properties with relatively small numerical deviations from the actual ones as demonstrated here.

The main idea is to use the ordinary kept/discarded strategy as outlined in Fig.~\ref{fig_3_kd} in conjunction with the log-gap discretization. States are thus discarded after each iteration and the energy scale separation is inevitably broken. Consequently, the eigenspectra are corrupted especially in the high energy sector as shown in Fig.~\ref{fig_app_spectra} for SC-AIM with $\varepsilon_d=-U/2$, $U=3\Delta$ and $\Gamma=\Delta$. Notably, $\varphi=0$ and $\varphi=\pi$ remain exact. 

In Fig.~\ref{fig_app_fast_abs}, we then demonstrate that the subgap portion of eigenenergies is nevertheless well usable at least in an approximate manner with a relative error of the order of just few percent when compared to the standard two-channel NRG. We stress however that such calculations might be dangerous without any reference data from exact Wilsonian approaches. Moreover, they should never be used for obtaining spectral functions due to the corruption of the high energy spectrum.

\vspace{3mm}

\section{Technical details of NRG implementations \label{sec_app_nrg} }

In the presented work, all model parameters are measured in units of $B$, where $2B$ is the width of the band. A typical value of the gap for SC-AIM as well as the gapped AIM with constant TDOS is set to $0.0005B$ if not stated otherwise. The effects of finite band width are thus essentially almost completely suppressed. Two standard NRG calculations have been implemented within the open source code of NRG Ljubljana: the standard two-channel NRG for SC-AIM \cite{Satori-1992} and the one-channel calculation for systems augment with metallic leads as described in Refs.~\cite{Diniz-2020,Zalom-2021}. The former has been performed at $\Lambda=4$ while for the later $\Lambda=2$ due to the one-channel nature of the calculation. In both cases at least $1000$ states have been kept.
 
The here developed log-gap NRG algorithm has been implemented in the Flexible DM-NRG Budapest code \cite{Toth-2008} by modifying its kept/discarded routines. Log-gap discretization \eqref{eq_log_gap_disc} was performed in a Mathematica script with a subsequent tridiagonalization performed in a stand-alone \textit{C++} code. The resulting Wilson chain parameters were then manually fed into the DM-NRG Budapest code. After the discarding step described of the diagonalization scheme, in Sec.~\ref{sec_alternating}, $500$ multiplets have been kept, which has been tested to be sufficient to have no impact on the sub-gap properties of the presented results. $\Lambda=2$ was set due to the one-channel nature of the Wilson chain.

To make the general idea of the log-gap NRG method more transparent and to streamline the presentation, the $z$-averaging was not employed. More specifically, the in-gap positions have been calculated at single value of $z=1$. This applies to the novel log-gap NRG approach introduced in this paper as well as to the all standard NRG calculations performed within NRG Ljubljana with the exception of supragap function shown in Fig.~\ref{fig_pascu}.


%


\end{document}